\documentclass[manuscript=letter,journal = nalefd]{achemso}

\usepackage[version=3]{mhchem} 
\usepackage{amssymb}

\usepackage[separate-uncertainty=true,multi-part-units=single]{siunitx}

\usepackage{tikz}
\usepackage{pgfplots}
\pgfplotsset{compat=newest} 
\usepackage[linktocpage=false, pdfstartpage=1, pdfstartview=FitV,colorlinks=false, linkcolor=false, citecolor=false, urlcolor=false, linktocpage=false,bookmarks=false,final, hidelinks]{hyperref}

\newcommand{\myautoref}[2][]{\hyperref[#2]{Figure~\ref*{#2}#1}}

\usepackage{verbatim}

\author{Mohamed-Raouf Amara}
\affiliation{%
 Laboratoire de Physique de l'\'Ecole Normale Sup\'erieure, ENS, Université PSL, CNRS, Sorbonne Universit\'e, Universit\'e Paris-Cité, F-75005 Paris, France
}%
\alsoaffiliation{%
 Division of Physics and Applied Physics, School of Physical and Mathematical Sciences, Nanyang Technological University, 637371, Singapore
}%

\author{Caixia Huo}%
\affiliation{%
 Division of Physics and Applied Physics, School of Physical and Mathematical Sciences, Nanyang Technological University, 637371, Singapore
}%
\alsoaffiliation{Institute of Materials/School of Materials Science and Engineering, Shanghai University, Shanghai 200444, China}
\alsoaffiliation{Shaoxing Institute of Technology, Shanghai University, Zhejiang 312000, China}
\author{Christophe Voisin}
\affiliation{%
 Laboratoire de Physique de l'\'Ecole Normale Sup\'erieure, ENS, Université PSL, CNRS, Sorbonne Universit\'e, Universit\'e Paris-Cité, F-75005 Paris, France
}%
\author{Qihua Xiong}
\affiliation{%
State Key Laboratory of Low-Dimensional Quantum Physics, Department of Physics, Tsinghua University, Beijing 100084, China.
}%
\alsoaffiliation{%
Frontier Science Center for Quantum Information, Beijing 100084, P.R. China
}%
\alsoaffiliation{%
Collaborative Innovation Center of Quantum Matter, Beijing 100084, P.R. China}%
\alsoaffiliation{%
Beijing Academy of Quantum Information Sciences, Beijing 100193, P.R. China}
\author{Carole Diederichs}
\email{carole.diederichs@phys.ens.fr}
\affiliation{%
 Laboratoire de Physique de l'\'Ecole Normale Sup\'erieure, ENS, Université PSL, CNRS, Sorbonne Universit\'e, Universit\'e Paris-Cité, F-75005 Paris, France
}%
\alsoaffiliation{%
 Institut Universitaire de France (IUF), 75231 Paris, France
}%
\title[Impact of bright-dark exciton thermal population mixing on the brightness of \ce{CsPbBr3} nanocrystals]
  {Impact of bright-dark exciton thermal population mixing on the brightness of \ce{CsPbBr3} nanocrystals}

\abbreviations{NC,LHP,PL,trPL}
\keywords{nanocrystals, perovskite, bright exciton, dark exciton, decay, exciton-phonon interaction}

\usepackage{style}
\myexternaldocument[SI]{supmat}

\begin{document}







\begin{abstract}
Understanding the interplay between bright and dark exciton states is crucial for deciphering the luminescence properties of low-dimensional materials.
	The origin of the outstanding brightness of lead halide perovskites remains elusive. %
	Here, we analyse temperature-dependent time-resolved photoluminescence to investigate the population mixing between bright and dark exciton sublevels in individual \ce{CsPbBr3} nanocrystals in the intermediate confinement regime. %
	We extract bright and dark exciton decay rates, and show quantitatively that the decay dynamics can only be reproduced with second-order phonon transitions. %
	Furthermore, we find that any exciton sublevel ordering is compatible with the most likely population transfer mechanism.
	The remarkable brightness of lead halide perovskite nanocrystals rather stems from a reduced asymmetry between bright-to-dark and dark-to-bright conversion originating from the peculiar second-order phonon-assisted transitions that freeze bright-dark conversion at low temperature together with the very fast radiative recombination and favourable degeneracy of the bright exciton state. %
\end{abstract}

Lead halide perovskites (LHPs) have attracted significant attention as radiation absorbers, in solar cells~\cite{greenEmergencePerovskiteSolar2014} and detectors~\cite{SunSingleCrystal2020}, and as light emitters, in light-emitting diodes~\cite{linPerovskiteLightemittingDiodes2018,fakharuddin2022perovskite} and lasers~\cite{sutherlandPerovskitePhotonicSources2016,zhu2015lead,zhangroomtemp2014}. In particular, low-dimensional all-inorganic LHPs have emerged in recent years as new emitters with remarkable optoelectronic properties, all the while being easily synthesized and processed~\cite{protesescuNanocrystalsCesiumLead2015a, wangAllInorganicColloidalPerovskite2015, sutherlandPerovskitePhotonicSources2016}. %
In their nanocrystal (NC) form, LHPs appear as promising candidates for quantum photonics applications as they exhibit a pure~\cite{huSuperiorOpticalProperties2015} and fast~\cite{rainoSingleCesiumLead2016} emission of single photons stable up to room temperature~\cite{parkRoomTemperatureSinglePhoton2015}. %
Low temperature studies have further evidenced a long optical coherence time~\cite{utzatCoherentSinglephotonEmission2019,lvQuantumInterferenceSingle2019,lvExcitonacousticPhononCoupling2021} as well as emission of correlated pairs of photons due to an efficient biexciton-exciton cascade at low temperature~\cite{tamaratDarkExcitonGround2020}, further asserting the quantum optical potential of these materials.

Nevertheless, some fundamental properties of LHP NCs remain to be understood. In particular, the origin of the brightness of LHPs, which is at the heart of their potential, is still debated. %
Taking into account the effect of a possible spatial inversion asymmetry and the large spin-orbit coupling in these heavy materials, it has been postulated that the Rashba effect may lead to a unique bright-dark exciton level order with a bright exciton ground state. While this effect would only be expected for the larger NCs~\cite{beckerBrightTripletExcitons2018,sercelExcitonFineStructure2019,sercelQuasicubicModelMetal2019}, it was proposed as a mechanism to explain the brightness of LHP NCs. %
However, Tamarat et al. demonstrated experimentally that the ground state is dark in quantum confined hybrid and all-inorganic perovskite NCs~\cite{fuUnravelingExcitonPhonon2018,tamaratGroundExcitonState2019,tamaratDarkExcitonGround2020}. %
For the prototypal \ce{CsPbBr3}, a dark ground state was only recently evidenced in bulk-like NCs~\cite{tamaratUniversalScalingLaws2023}, in contrast with the predictions involving the Rashba effect~\cite{sercelExcitonFineStructure2019,sercelQuasicubicModelMetal2019}.
In these studies on quantum confined NCs, an alternative explanation for the brightness of LHPs was proposed which does not require a peculiar bright-dark level inversion. It was rather suggested that bright-dark thermal population mixing was greatly reduced in LHPs with vanishing first-order one-phonon transitions. %

In this work, we analyse temperature-dependent time-resolved photoluminescence experiments on individual \ce{CsPbBr3} NCs in the intermediate confinement regime. %
We evidence thermal population transfer between the bright and dark exciton states. %
Taking into account both first- and second-order phonon-assisted population transfer mechanisms, we model bright-dark population dynamics. We find that first-order transitions are incompatible with our observations, and that two types of second-order processes can reproduce the dynamics. %
Due to the large energy splittings in quantum confined \ce{CsPbBr3} NCs, the most probable mechanism is a second-order process where the energy difference between the two phonon modes involved matches the bright-dark energy splitting. %
As a consequence, we rationalise that the exciton sublevels ordering is irrelevant to the brightness of lead halide perovskites which can readily be explained by a combination of favourable factors that lead to a majority bright exciton recombination at all temperatures. %

\begin{figure}
	\centering
	\includegraphics{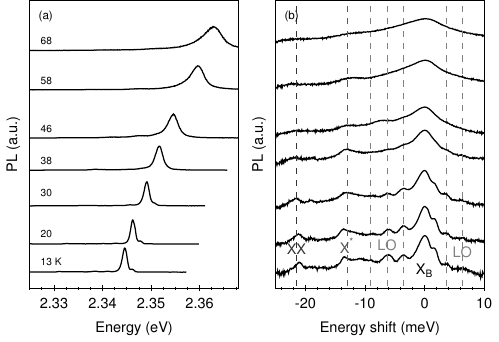}
	\caption{Single NC spectrum as a function of temperature displaying the emission of the bright triplet exciton (X\textsubscript{B}), optical phonon replica (LO), the charged exciton (X\textsuperscript{*}) and the biexciton (XX). Same data in (a) linear scale and (b) logarithmic scale with all spectra horizontally centered at the lowest bright exciton zero phonon line.}
	\label{fig:singleNCs}
\end{figure}

%
We studied the temperature-dependence of the steady-state and time-resolved photoluminescence from 4.5 up to $\sim$\SI{100}{\kelvin}. %
As previously reported~\cite{fuNeutralChargedExciton2017,beckerBrightTripletExcitons2018,choExcitonPhononTrion2022,amaraSpectralFingerprintQuantum2023}, typical single NC spectra consist in two or three main peaks attributed to the bright triplet exciton together with Stokes-shifted peaks related to the charged exciton, the biexciton and their respective optical phonon replica. \myautoref{fig:singleNCs} shows the example of a NC with two bright exciton peaks. %
With increasing temperature, emission lines exhibit a global blue-shift with an average slope of~\SI{0.3}{\milli\electronvolt\per\kelvin} excluding any phase transition (see note~\ref{SIsec:allenergies}). %
This is similar to the slopes determined in both hybrid~\cite{fuUnravelingExcitonPhonon2018,tamaratGroundExcitonState2019} and all-inorganic LHP NCs~\cite{ramadeExcitonphononCouplingCsPbBr32018} and slightly smaller than the slope found in bulk \ce{CsPbBr3}~\cite{guoDynamicEmissionStokes2019}. %
This linear increase of emission energy with temperature is at contrast with that observed for wurtzite and zinc-blende semiconductors where bandgaps decrease with increasing temperature. %
This bandgap increase with temperature, which has already been evidenced in lead-based semiconductors~\cite{gaponenkoTemperaturedependentPhotoluminescencePbS2010} and LHPs~\cite{sebastianExcitonicEmissionsAbovebandgap2015,ramadeFineStructureExcitons2018a}, points towards a peculiar electron-phonon interaction.

%
Increasing temperature also leads to a broadening of the emission lines, the study of which is challenging as the entire emission manifold is within a \SI{50}{\milli\electronvolt} bandwidth. %
Exciton dephasing via Fröhlich interaction was previously evidenced, both in ensembles~\cite{wrightElectronPhononCoupling2016,dirollLowTemperatureAbsorptionPhotoluminescence2018} and at the single object level~\cite{fuUnravelingExcitonPhonon2018,ramadeExcitonphononCouplingCsPbBr32018}, revealing an effective optical phonon mode energy close to a bulk mode at~\SI{16}{\milli\electronvolt}. %
In these studies, exciton dephasing related to acoustic phonons was found to be negligible. %
In our measurements, the exciton linewidth shows a similar increase with temperature with a low-temperature linewidth already on the order of \SI{1}{\milli\electronvolt} suggesting that the influence of acoustic-phonon coupling is likely underestimated in the low-temperature linewidths. %
Indeed, resonant photoluminescence of single \ce{CsPbI3} NCs~\cite{lvExcitonacousticPhononCoupling2021} has revealed the role of acoustic phonons in the exciton dephasing process with an overwhelming contribution of acoustic phonons. %
Besides, as investigated theoretically in \ce{CsPbBr3} NCs, exciton-phonon coupling is expected to be dominated by different optical phonon modes depending on the NC size and surface passivation~\cite{rainoUltranarrowRoomtemperatureEmission2022}.
Detailed examination of the exciton linewidth broadening and assignment to specific optical phonon modes thus remains elusive at this stage. %
%

Further insight into the electronic structure of these NCs can be sought from the fast decay times, contrasting with other conventional materials. %
At all temperatures, the time-resolved PL of the bright exciton in single \ce{CsPbBr3} NCs is well fitted by a bi-exponential decay with a short sub-\SI{100}{\pico\second} decay time and a longer \si{\nano\second} decay time. %
\begin{figure*}
	\centering
		\includegraphics[width=\textwidth]{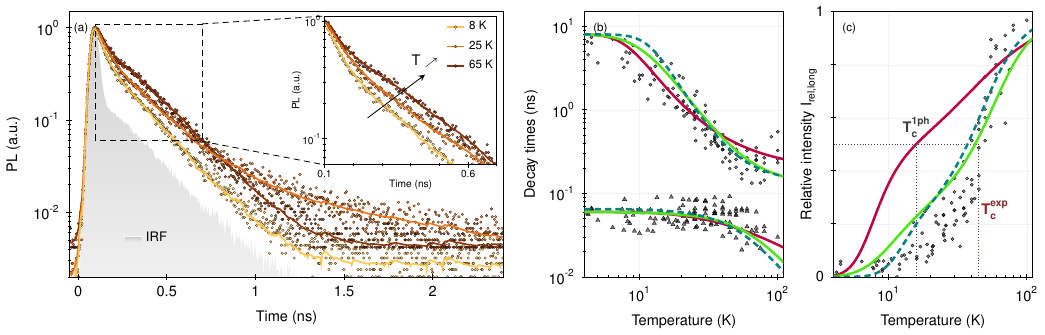}
	\caption{Decay dynamics as a function of temperature. (a) Time-resolved photoluminescence of a single \ce{CsPbBr3} NC at 8, 25 and \SI{65}{\kelvin}. (b,c) Summary of decay dynamics measured on several single NCs. (b) Long and short decay times and (c) long decay component fractional intensity. Experimental data are obtained by fitting the transients at each temperature by a bi-exponential decay and are fitted by the models described in the main text. Colour codes for the fits are the same as in \myautoref[]{fig:BoDNphmaps}: first-order in solid red, second-order sum process in solid green and second-order difference process in dashed green.}
	\label{fig:singleNCdecaysonly}
\end{figure*}
At \SI{4.5}{\kelvin}, the emission of the bright triplet exciton happens mainly through the sub-\SI{100}{\pico\second} decay channel (\myautoref{fig:singleNCdecaysonly} and \myautoref{SIfig:decaystats}). %
As temperature is increased (\myautoref[a]{fig:singleNCdecaysonly} and \myautoref[b]{SIfig:dataset}), PL decays display a behaviour characteristic of bright-dark exciton thermal population mixing with the gradual vanishing of the short-decay channel together with the emergence of a long decay component that shortens, gains weight and becomes the dominant decay channel at higher temperatures ($T>T_c^{\text{exp}}\sim\SI{45}{\kelvin}$). %
Examples of fitted decays are shown in \myautoref[a]{fig:singleNCdecaysonly} and \myautoref[b]{SIfig:dataset} for two individual NCs while the experimental results obtained on the whole set of NCs studied are summarised in \myautoref[b,c]{fig:singleNCdecaysonly}. %
%

We model our experiments with a four level system comprised of: a ground state (G), an excited state (E), one bright exciton state (B), accounting for the triplet, and one dark (D) exciton state, with respective decay rates $\Gamma_\text{B}$ and $\Gamma_\text{D}$ (\myautoref[a]{fig:BoDNphmaps}). %
Considering that the non-resonant excitation used in our experiments likely yields an equal population in the bright triplet and dark singlet states, we set the fraction of the population in the bright state after relaxation from the higher lying state (E) to ${a=3/4}$ owing to its degeneracy. %
In our experiments, we measure solely the bright triplet exciton decay which is proportional to the bright exciton state population. %

Within this model, bright-dark transitions, with rates $\gamma_{\uparrow\,\downarrow}$, are modelled as thermally-activated phonon-assisted processes. %
In other nanoemitters, such as II-VI NCs~\cite{labeauTemperatureDependenceLuminescence2003,werschlerCouplingExcitonsDiscrete2016} or carbon nanotubes~\cite{bergerTemperatureDependenceExciton2007,gokusMonoBiexponentialLuminescence2010}, phonon-driven transitions between bright and dark exciton states have been attributed to acoustic phonons whose energy matches the bright-dark splitting. %
The upward and downward transition rates are then ${\gamma_\uparrow = \gamma_0 N}$ and ${\gamma_\downarrow = \gamma_0\left(1+N\right)}$, with $N$ the Bose-Einstein occupation factor and $\gamma_0$ the transition rate constant. %
For LHPs, although the same model has been employed~\cite{chenCompositionDependentEnergySplitting2018,rossiIntenseDarkExciton2020,rossiSizedependentDarkExciton2020}, interpreting in terms of a single acoustic phonon mode does not hold due to bright-dark energy splittings ($\gtrsim$~\SI{5}{\milli\electronvolt}) surpassing available acoustic phonon energies $\sim$~\SI{1}{\milli\electronvolt}. %
Moreover, in the reference cubic perovskite structure, polar exciton-acoustic phonon coupling is expected to be forbidden~\cite{evenCarrierScatteringProcesses2016}, while coupling via deformation potentials is inhibited by lattice anharmonicity~\cite{miyataLeadHalidePerovskites2017}. Consequently, exciton-phonon interaction is expected primarily via Fröhlich interaction.

Here, we consider both one- and two-optical phonon transitions and investigate the bi-exponential decay evolution with temperature. The decay dynamics can be calculated using four parameters ($\Gamma_\text{B}, \Gamma_\text{D}, \gamma_0, \Delta_\text{BD}$), where $\gamma_0$ is the transition rate constant and $\Delta_\text{BD}$ the bright-dark energy splitting. Comparing the low- and high-temperature asymptotic limits to our experimental results (see note~\ref{SIsec:simparams}), we constrain $\Gamma_\text{B}$ and $\Gamma_\text{D}$, leaving only two adjustable parameters: $\gamma_0$ and $\Delta_\text{BD}$.
The decay dynamics are then simulated across several orders of magnitude in the $(\gamma_0,\ \Delta_\text{BD})$ phase space, with ${\SI{0.01}{\nano\second^{-1}}<\gamma_0<\SI{100}{\nano\second^{-1}}}$ and ${\SI{0.01}{\milli\electronvolt}<\Delta_\text{BD}<\SI{100}{\milli\electronvolt}}$, and several phonon-assisted mixing scenarios as well bright-dark orderings are investigated.

\begin{figure}
	\includegraphics{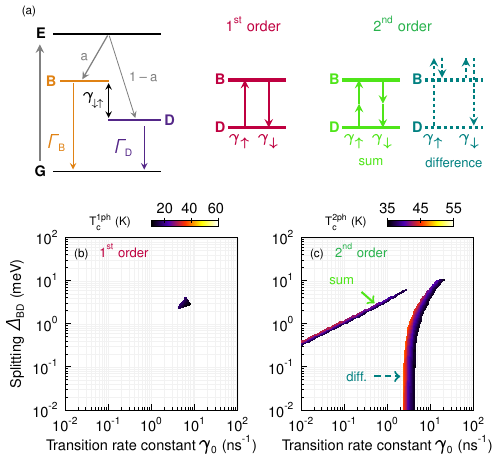}
	\caption{Phase space mapping of first- and second-order transitions. (a) Schematic of the system comprised of a ground state (G), an excited state (E), a bright exciton state (B) and a dark exciton state (D) and the coupling rates of the model. The three bright-dark thermal population mixing schemes explored are detailed. (b,c) Characteristic temperature $T_c^{\text{1ph}}$, resp. $T_c^{\text{2ph}}$, in the phase subspace defined by the intersection of experimental requirements for (b) first-order transitions and (c) second-order transitions respectively.}
	\label{fig:BoDNphmaps}
\end{figure}

As detailed in note~\ref{SIsec:procedure}, we calculate the long-decay-time relative intensity in the low- and high-temperature limits, $I_\text{rel,long}^\text{0 K}$ and $I_\text{rel,long}^\text{100 K}$, as well as the long decay time in the high temperature limit, $\tau_{\text{long}}^{100 K}$, comparing them to our experimental results. %
For each quantity, we constrain the phase-space to the $(\gamma_0,\, \Delta_\text{BD})$ that yield dynamics consistent with our measurements. %
As these criteria define the range within which all experimental decays are observed, we finally take their intersection. %
For one-phonon transitions (\myautoref[b]{fig:BoDNphmaps}), we identify a limited region around $\gamma_0 \sim \SI{6}{\nano\second^{-1}}$ and $\Delta_\text{BD} \sim \SI{3}{\milli\electronvolt}$ which satisfies these requirements. %
While this energy splitting $\Delta_\text{BD}$ is close to the expected bright-dark splittings in bulk \ce{CsPbBr3}~\cite{tamaratDarkExcitonGround2020,tamaratUniversalScalingLaws2023}, the characteristic thermal mixing temperature $T_c^{\text{1ph}}$ (defined as the temperature at which the short and long component intensities are equal) is much lower than observed, i.e. $T_c^{\text{1ph}}\sim \SI{15\pm 5}{\kelvin}$ while $T_c^{\text{exp}}\sim\SI{45\pm 5}{\kelvin}$. %
This is exemplified in \myautoref[b,c]{fig:singleNCdecaysonly} where the decay times dynamics can be reasonably well reproduced while the relative intensity of the decay components dynamics cannot. %
This demonstrates that first-order phonon-assisted mixing cannot be the dominant mechanism in our system. %
As an alternative, a second-order process has been proposed for hybrid and all-inorganic perovskite NCs~\cite{fuUnravelingExcitonPhonon2018,tamaratGroundExcitonState2019,tamaratDarkExcitonGround2020} where bright-dark energy splittings, varying over an order of magnitude from $\sim$~\SI{0.3}{\milli\electronvolt} to $\sim$~\SI{3}{\milli\electronvolt} depending on the size and actual composition, could be matched by the energy difference between two optical phonons.

Following the same procedure as for first-order exciton-phonon coupling, we determine the characteristic mixing temperature for second-order coupling and investigate the two types of two-phonon processes sketched in \myautoref[a]{fig:BoDNphmaps}. %
The two-phonon difference process, involving phonon modes $i$ and $j$ with energies $\hbar\omega_i$, resp. $\hbar\omega_j$, verifies $\hbar(\omega_j-\omega_i) = \Delta_\text{BD}$ with $\omega_j>\omega_i$ such that both the upward $\gamma_\uparrow = \gamma_0 N_j \left(N_i+1\right)$ and downward $\gamma_\downarrow = \gamma_0 N_i \left(N_j+1\right)$ rates model the simultaneous emission and absorption of phonons. %
On the other hand, the other second-order process investigated, the sum process, verifies $\hbar(\omega_i+\omega_j) = \Delta_\text{BD}$. %
The upward rate $\gamma_\uparrow=\gamma_0N_iN_j$ thus results from the absorption of two phonons while the downward rate results from the emission of two phonons $\gamma_{\downarrow}=\gamma_0\left( 1+N_i\right)\left( 1+N_j\right)$. %

Analysis of the phase spaces in \myautoref[c]{fig:BoDNphmaps} reveals that both second-order mixing schemes can reproduce the decay dynamics. Examples of simulated decay dynamics are shown in \myautoref[b,c]{fig:singleNCdecaysonly} together with the experimental data. %
%
We highlight that for both second-order mixing models the actual phase spaces are much larger than shown in \myautoref[c]{fig:BoDNphmaps} as the energies of the two phonons can further be varied ensuring respectively $\hbar(\omega_2 - \omega_1)=\Delta_\text{BD}$ and $\hbar(\omega_1 + \omega_2)=\Delta_\text{BD}$. Here, we have assumed $\hbar\omega_1 = \SI{3}{\milli\electronvolt}$ for the difference process, matching the lowest optical phonon energy, and $\hbar\omega_1 = \hbar\omega_2 = \Delta_\text{BD}/2$ for the sum process. %
Alternative possibilities explored in \myautoref{SIfig:CoT_extra} by varying the lowest phonon mode energy yield similar phase spaces for both second-order mixing schemes. %
In all cases, a range of $\{\Delta_\text{BD},\,\gamma_0\}$ couples that verify the criteria extracted from experiments and yield $T_c^{\text{2ph}}\sim T_c^{\text{exp}}$ is found (\myautoref[c]{fig:BoDNphmaps}).

For comprehensiveness, we also considered the case of a bright state below the dark state. %
In this case, only the difference process can reproduce the decay dynamics as it is insensitive to the actual level ordering. %
This suggests that the actual level ordering is not relevant to reproduce the decay dynamics. %
In contrast, the necessity of second-order phonon-assisted processes for population mixing is crucial to reproduce the dynamics. %
Based on theoretical predictions including the Rashba effect~\cite{sercelQuasicubicModelMetal2019,sercelExcitonFineStructure2019} and recent results~\cite{tamaratUniversalScalingLaws2023}, the scenario of a bright ground state is ruled out for the weakly quantum-confined NCs investigated here. %
Consequently, at this stage, with a dark ground state, the observed dynamics are reproduced using second-order transitions regardless of the specific process, be it the sum or the difference process.%
%

Magneto-optical spectroscopy of bulk-like \ce{CsPbBr3} NCs unveiled a conventional fine structure with a dark state below the bright triplet and a bulk bright-dark energy splitting close to \SI{4}{\milli\electronvolt}~\cite{tamaratUniversalScalingLaws2023}. %
Without a Rashba effect induced level inversion, confinement is expected to lead to an increase of bright-dark energy splittings reaching up to $\sim$~\SI{10}{\milli\electronvolt} for NCs in the intermediate confinement regime~\cite{sercelQuasicubicModelMetal2019,sercelExcitonFineStructure2019,benaichMultibandModelTetragonal2020}. %
Our magneto-optical spectroscopy on single \ce{CsPbBr3} NCs in the intermediate confinement regime reveals a weak emergence of a dark singlet state beneath the bright triplet. %
At the highest field strengths, we observe the emergence of a peak below the bright triplet at $\sim~12$ and $\sim~\SI{14}{\milli\electronvolt}$ for the NC emitting at 2.42 and \SI{2.48}{\electronvolt} respectively  (see note~\ref{SIsec:darkexp} and \myautoref{SIfig:PLfBB}). %
These bright-dark energy splittings compare well with the recently determined scaling law for LHP NCs, falling within the range of normalised splittings reported for \ce{CsPbI3} NCs in a similar confinement regime~\cite{tamaratUniversalScalingLaws2023}. %
Unfortunately temperature-dependent photoluminescence decay measurements were not conducted on these two NCs, precluding the use of these energy splitting to model the dynamics and derive a definite transition rate constant. %
%
Nonetheless, these results allow us to estimate the transition rates based on our simulations and the general decay dynamics reported here. %
Notably, for such large energy splittings, only the difference two-phonon process can account for the observed dynamics with corresponding transition rates on the order of \SI{10}{\nano\second^{-1}} for an energy splitting of $\sim~\SI{13}{\milli\electronvolt}$. This aligns closely with a resonance of the two-phonon process involving phonon modes $E_1 = \SI{6}{\milli\electronvolt}$ and $E_2 = \SI{19}{\milli\electronvolt}$ which were previously identified~\cite{choExcitonPhononTrion2022,amaraSpectralFingerprintQuantum2023}. %
%

Before investigating the impact of this peculiar bright-dark exciton thermal population mixing, we revisit the decay dynamics briefly. In addition to the earlier discussed low-temperature bright-dark dynamics, we note a remarkable onset of an increase of the long decay time around \SI{80}{\kelvin} (\myautoref[b]{fig:singleNCdecaysonly}). %
This contrasts with the usual activation of additional non-radiative decay pathways with increasing temperature resulting in reduced decay times. %
Lifetime increase with temperature has previously been reported in LHPs both in ensembles~\cite{weiTemperaturedependentExcitonicPhotoluminescence2016,dirollLowTemperatureAbsorptionPhotoluminescence2018} and at the single-object level~\cite{liuCationEffectExcitons2019,rossiIntenseDarkExciton2020}, with a bright exciton decay time shorter at low temperature than at room temperature and exhibiting a singular increase at intermediate temperatures. %
This phenomenon cannot be solely attributed to the bright-dark thermal population transfer as it yields monotonous variation of the decay time. This suggests the involvement of another process at higher temperatures. %
Similar behaviour has been observed in various weakly-confined systems~\cite{itohSizedependentRadiativeDecay1990,misawaSuperradianceQuenchingConfined1991,ithurriaColloidalNanoplateletsTwodimensional2011,morganExcitonLocalizationRadiative2019} and attributed to a loss of oscillator strength \emph{i.e.} a reduced exciton coherence volume induced by thermally activated phonon interactions~\cite{misawaSuperradianceQuenchingConfined1991,bellessaQuantumsizeEffectsRadiative1998,morganExcitonLocalizationRadiative2019}. %
Interestingly, the threshold temperature for the lifetime increase observed in our experiments ($\sim$~\SI{80}{\kelvin}) corresponds to a thermal energy of \SI{6.9}{\milli\electronvolt} close to the reported optical phonon mode at \SI{6.3}{\milli\electronvolt}~\cite{fuNeutralChargedExciton2017,choExcitonPhononTrion2022,amaraSpectralFingerprintQuantum2023}, suggesting a significant influence of this optical phonon mode on the thermally-induced loss of exciton coherence. %
Thus, since such coherence effects manifest only in the radiative lifetime, the observed increase of the bright exciton decay time strongly suggests a predominantly radiative recombination of the bright exciton.

\begin{figure}[htbp]
    \centering
    \includegraphics{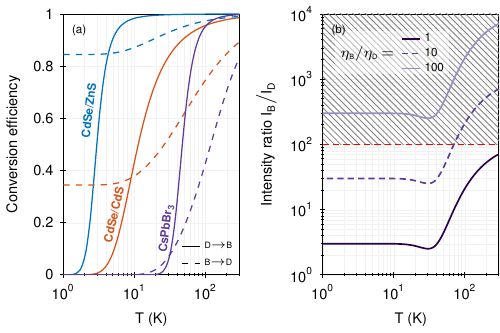}
    \caption{Bright-dark conversion efficiency and relative intensity. (a) Conversion efficiency between bright and dark exciton in several materials. Transition rates for \ce{CsPbBr3} are $\gamma_0=\SI{10}{\nano\second^{-1}}$, $\Gamma_\text{B}=\SI{15}{\nano\second^{-1}}$, $\Gamma_\text{D}=\SI{0.12}{\nano\second^{-1}}$ with phonon energies $E_{\text{ph},1}=\SI{6}{\milli\electronvolt}$, $E_{\text{ph},2}=\SI{19}{\milli\electronvolt}$ and $a=0.75$ while parameters for CdSe/ZnS and CdSe/CdS are extracted respectively from~\cite{biadalaDirectObservationTwo2009} and~\cite{werschlerCouplingExcitonsDiscrete2016}. (b) Intensity ratio between the bright and dark state emission in \ce{CsPbBr3} for several values of $\eta_\text{B}/\eta_\text{D}$ using the same parameters as in (a). The dashed line shows $I_\text{B}/I_\text{D}=100$ corresponding to the upper bound for a detectable dark exciton emission in our experiments while the dashed region corresponds to an undetectable dark exciton emission.}
    \label{fig:IBID}
\end{figure}

We now compare the temperature-dependent transition rates to $\Gamma_\text{B}$ and $\Gamma_\text{D}$ to examine the brightness of \ce{CsPbBr3} NCs in the absence of a magnetic field. %
First, we investigate the conversion efficiency, i.e. the fraction of a state population converting to the other exciton state (\myautoref[a]{fig:IBID} and note~\ref{SIsec:conversionefficiency}). %
The conversion efficiency from bright to dark exciton, resp. dark to bright exciton levels, is defined as $\gamma_\downarrow/(\Gamma_\text{B} + \gamma_\downarrow)$, resp. $\gamma_\uparrow/(\Gamma_\text{D} + \gamma_\uparrow)$.
In CdSe-based NCs, bright excitons are efficiently converted into dark excitons at low temperatures while dark-bright conversion is frozen. With increasing temperature, conversion from dark to bright exciton becomes possible and even more efficient than bright-to-dark conversion from a few Kelvin onwards. %
In contrast, for LHP NCs, both bright-dark and dark-bright exciton conversions are frozen at low temperature (see note~\ref{SIsec:conversionefficiency}). As temperature is increased, both conversion efficiencies increase and dark-bright conversion becomes more efficient than bright-dark conversion. %
As such, the asymmetry between dark-bright and bright-dark exciton conversion is less pronounced in LHP NCs, which, coupled with the favourable degeneracy of the bright state, results in a larger overall population in the bright exciton state (\myautoref[b]{fig:IBID}).

Secondly, we evaluate the relative intensity of bright and dark exciton emission (see note~\ref{SIsec:intint}). %
For CdSe-based NCs, where the bright and dark exciton states have similar quantum yields $\eta_\text{B}=\eta_\text{D}$, bright exciton emission is expected to be significantly weaker than dark exciton emission at the lowest temperatures ($I_\text{B}/I_\text{D}\sim0.1$) due to comparable $\gamma_0$ and $\Gamma_\text{B}$, consistent with experimental observations~\cite{biadalaDirectObservationTwo2009,werschlerCouplingExcitonsDiscrete2016}. %
For LHP NCs however, as shown in~\myautoref[b]{fig:IBID}, bright exciton emission is expected to be stronger than dark exciton emission at the lowest temperatures ($I_\text{B}/I_\text{D}\rightarrow3\eta_\text{B}/\eta_\text{D}$). %
This asymptotical behaviour is independent of the actual rates within the system and depends only on the relative quantum yields of the bright and dark exciton states. %
Note that this holds true regardless of the actual exciton sublevels ordering (see note~\ref{SIsec:intint}). %
The absence of dark state emission in intermediately confined \ce{CsPbBr3} NCs, be it in this work or others~\cite{cannesonNegativelyChargedDark2017,fuNeutralChargedExciton2017,rossiIntenseDarkExciton2020,choExcitonPhononTrion2022,amaraSpectralFingerprintQuantum2023}, thus gives direct information on the relative quantum yields of the bright and dark exciton state. %
Considering our experimental signal-to-noise ratio, we expect that intensity ratios $I_\text{B}/I_\text{D}\leq 100$ should have been detected. The absence of dark state emission allows us to constrain the ratio of quantum yields $\eta_\text{B}/\eta_\text{D}$. %
As shown in~\myautoref[b]{fig:IBID}, a $\eta_\text{B}/\eta_\text{D}=10$ ratio would result in observable dark exciton state emission below \SI{70}{\kelvin}, whereas with a ratio $\eta_\text{B}/\eta_\text{D}=100$ the dark exciton state would remain unobserved at all temperatures. %
Numerically, we find that the quantum yields ratio of at least 40 is necessary for the dark state emission to remain unobserved at all temperatures, \emph{i.e.} for $I_\text{B}/I_\text{D}\geq 100$ at all temperatures. %
Furthermore, considering the output from each state, regardless of its radiative or non-radiative character, which is equivalent to setting $\eta_B=\eta_D$ in~\myautoref[b]{fig:IBID}, reveals that a majority of excitons recombine from the bright state.
Thus, regardless of the radiative or non-radiative character of the dark exciton recombination, the emission of LHPs remains bright because a majority of excitons recombine from the bright exciton state regardless of the temperature. %

Through temperature-dependent and time-resolved PL measurements on single \ce{CsPbBr3} NCs in the intermediate confinement regime, we find that the bright exciton decay dynamics are driven by second-order bright/dark exciton thermal population transfer at low temperature and suggest a dominant contribution of exciton-phonon dephasing at higher temperature. %
We ruled out the ability of one-phonon transitions to reproduce the observed decay dynamics and emphasised that while a one-phonon mixing model may reproduce the decay times evolution with temperature, it cannot concurrently reproduce that of the decay components intensities. %
This is a remarkable feature of intermediately confined LHPs where one-phonon bright-dark transitions are suppressed and bright-dark population mixing is slowed down. %
Notably, we find here that two types of two-phonon processes, the sum- and the difference-process, could reproduce the decay dynamics in single \ce{CsPbBr3} NCs with a bright-dark energy splitting in the few \si{\milli\electronvolt} range. %
However, for larger energy splittings, only the difference process can reproduce the decay dynamics with a corresponding transition rate constant on the order of \SI{10}{\nano\second^{-1}}. %
In addition, we find that the actual bright-dark ordering has a marginal impact on the overall dynamics and brightness. %
Our study thus provides direct insights into the distinct features of bright-dark dynamics in intermediately confined \ce{CsPbBr3} NCs in which the emission happens mainly radiatively through the bright exciton. This is attributed to (i) the weak efficiency of the bright-dark thermal population mixing at low temperature, (ii) favourable transition rates with a fast radiative recombination of the bright exciton and a slow recombination of the dark exciton, and (iii) a degeneracy of exciton levels favouring the bright state. %


\section{Funding}
This work was supported by the French National Research Agency (ANR) through the project IPER-Nano2 (ANR-18-CE30-0023). 
Q.X. gratefully acknowledges strong funding support by the National Key Research and Development Program of China (grant No. 2022YFA1204700), National Natural Science Foundation of China (grant No. 12020101003 and 12250710126), funding support from the State Key Laboratory of Low-Dimensional Quantum Physics of Tsinghua University and the Tsinghua University Initiative Scientific Research Program.

\begin{suppinfo}
Experimental methods, time-resolved experiments analysis method, bright-dark dynamics model, and supporting data
\end{suppinfo}

\begin{acknowledgement}
The authors thank Yannick Chassagneux and Robson Ferreira at LPENS for fruitful discussions.
\end{acknowledgement}

\bibliography{bdmanu}


\end{document}


\title{\texorpdfstring{Supporting information:\\ Impact of bright-dark exciton thermal population mixing on the brightness of CsPbBr\textsubscript{3} nanocrystals}{Supplementary information: Impact of bright-dark exciton thermal population mixing on the brightness of CsPbBr\textsubscript{3} nanocrystals}}%

\author{Mohamed-Raouf Amara}
\affiliation{%
 Laboratoire de Physique de l'\'Ecole Normale Sup\'erieure, ENS, Université PSL, CNRS, Sorbonne Universit\'e, Universit\'e Paris-Cité, F-75005 Paris, France
}%
\affiliation{%
 Division of Physics and Applied Physics, School of Physical and Mathematical Sciences, Nanyang Technological University, 637371, Singapore
}%

\author{Caixia Huo}%
\affiliation{%
 Division of Physics and Applied Physics, School of Physical and Mathematical Sciences, Nanyang Technological University, 637371, Singapore
}%
\affiliation{Institute of Materials/School of Materials Science and Engineering, Shanghai University, Shanghai 200444, China}
\affiliation{Shaoxing Institute of Technology, Shanghai University, Zhejiang 312000, China}
\author{Christophe Voisin}
\affiliation{%
 Laboratoire de Physique de l'\'Ecole Normale Sup\'erieure, ENS, Université PSL, CNRS, Sorbonne Universit\'e, Universit\'e Paris-Cité, F-75005 Paris, France
}%
\author{Qihua Xiong}
\affiliation{%
State Key Laboratory of Low-Dimensional Quantum Physics, Department of Physics, Tsinghua University, Beijing 100084, China.
}%
\affiliation{%
Frontier Science Center for Quantum Information, Beijing 100084, P.R. China
}%
\affiliation{%
Collaborative Innovation Center of Quantum Matter, Beijing 100084, P.R. China}%
\affiliation{%
Beijing Academy of Quantum Information Sciences, Beijing 100193, P.R. China}
\author{Carole Diederichs}
\email{Corresponding author: carole.diederichs@phys.ens.fr}
\affiliation{%
 Laboratoire de Physique de l'\'Ecole Normale Sup\'erieure, ENS, Université PSL, CNRS, Sorbonne Universit\'e, Universit\'e Paris-Cité, F-75005 Paris, France
}%
\affiliation{%
 Institut Universitaire de France (IUF), 75231 Paris, France
}%

\maketitle

\setcounter{equation}{0}
\setcounter{section}{0}
\setcounter{figure}{0}
\setcounter{table}{0}
\setcounter{page}{1}
\renewcommand{\thefigure}{S\arabic{figure}}
\renewcommand{\thetable}{S\arabic{table}}
\renewcommand{\thesection}{S\Roman{section}}
\renewcommand{\bibnumfmt}[1]{[S#1]}
\renewcommand{\citenumfont}[1]{S#1}
\renewcommand{\theequation}{S\arabic{equation}}

\section{Experimental aspects}

\subsection{Synthesis and characterisation}

\begin{figure}[h]
    \centering
	\includegraphics{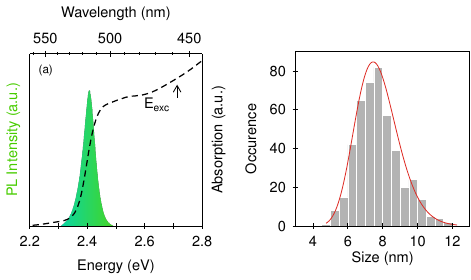}
    \caption{Synthesis characterisation. (a) Ensemble room-temperature photoluminescence and absorption spectra. The excitation energy is shown as $E_\text{exc}$. (b) Size-distribution histogram obtained from transmission electron microscopy.
	}
    \label{fig:charac}
\end{figure}

Cesium lead bromide nanocrystals (\ch{CsPbBr3} NCs) were grown using a modified previously described method~\cite{beckerBrightTripletExcitons2018}. Hereafter, we highlight only the main differences.
The injection temperature is lowered to \SI{180}{\celsius} to yield smaller NCs. After injection, the reaction was stopped briefly in an ice water bath and then left at room temperature. The solution is then purified by several rounds of centrifugation and re-dispersion in hexane until the obtained solution remains clear and stable. The solution is then filtered using a \SI{20}{\nano\meter} membrane filter to remove any by-products of the synthesis that might remain after the purification process. %
Transmission electron microscopy was performed on the synthesized NCs and revealed the presence of nearly cubic NCs (average aspect ratio $\num{0.89}\pm\num{0.09}$) with a mean edge length of \SI{7.7\pm1.2}{\nano\meter} and sizes ranging from \SIrange{4}{12}{\nano\meter}. Detailed information about the size, morphology and transmission electron microscopy images of the synthesised NCs can be found in ref.~\cite{amaraSpectralFingerprintQuantum2023} under the label \emph{synthesis A}. 

\begin{figure*}
	\sffamily
	\includegraphics{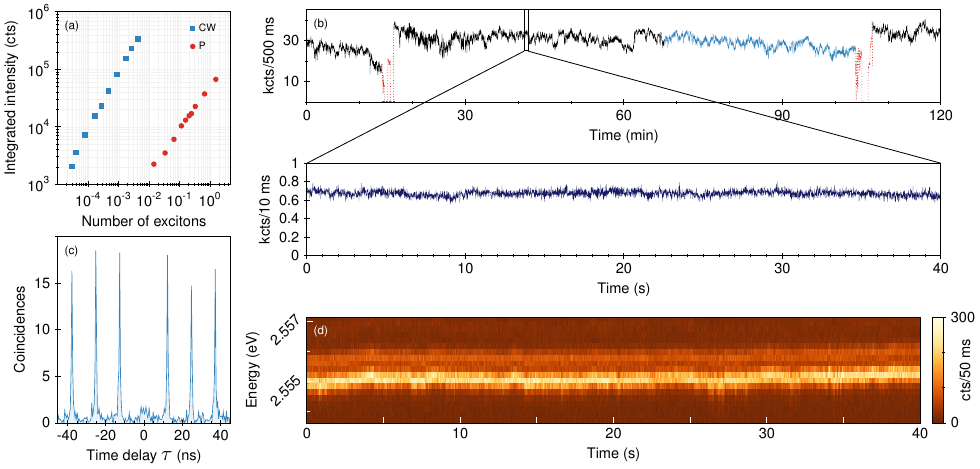}
	\caption{Additional experimental aspects at \SI{4.5}{\kelvin}. (a) Representative excitation power density dependence of the integrated emission intensity of a single NC under continuous wave (CW) and pulsed excitation. (b) PL intensity of a single NC under CW excitation revealing the stability of the emission over the course of hours. Time intervals shown in red correspond to the reoptimisation of the position of the excitation beam and focal, and the time interval highlighted in light blue shows a continuous small decrease of the PL intensity due to the slight drift of the sample.  (c) Raw coincidence counts of second-order correlation measurements performed at \SI{4.5}{\kelvin} showing photon antibunching at zero time delay. (d) PL time trace showing the minimal spectral diffusion observed at our resolution.}
	\label{fgr:exppow}
\end{figure*}

\subsection{Sample preparation}

For single NC measurements, the solution was diluted $\sim$~\num{2000}$\times$ in a 2\% solution of polystyrene in toluene. The targeted concentration was achieved by multiple rounds of dilution for accuracy/reproducibility. %
The solution was then spin-coated on either a glass or a distributed Bragg reflector (DBR) substrate to increase the collection efficiency. %

\subsection{Optical characterisation}

For single particle measurements, samples were mounted on the cold finger of a continuous liquid-helium flow cryostat cooled down to \SI{4.5}{\kelvin}. %
All optical measurements were performed with a home-built scanning micro-PL setup. Excitation and detection of the PL emission were realised with the same 50$\times$ long-working-distance dry microscope objective with NA=0.65. %
NCs were excited either by a continuous diode laser at \SI{457}{\nano\meter} or a pulsed \si{\femto\second} frequency-doubled Ti:Sa laser (repetition rate: \SI{80}{\mega\hertz}) tunable between \SI{400}{\nano\meter} and \SI{520}{\nano\meter}. %
The PL emission was dispersed with a diffraction grating and detected with a charge-coupled-device (CCD) camera. %
For time-resolved measurements, the PL emission was directed via the second output of the monochromator to two single-photon avalanche photodiodes (Micro Photon Devices SPADs) with a nominal dark count rate of less than \SI{50}{cts\per\second}. The overall resolution of the time-resolved setup is further discussed in the following. %
This setup layout enabled us to spectrally resolve the emission of single \ch{CsPbBr3} NCs with both the CCD and APDs.

Despite the low excitation power density kept at all times, charged exciton and biexciton emission is observed on the majority of NCs. \myautoref[a]{fgr:exppow} shows a representative power density dependence of the emission of a single NC displayed as a function of the average number of excitons generated respectively per lifetime and per pulse for continuous and pulsed excitation, assuming an absorption cross section $\sigma \sim \SI{4e-14}{\centi\meter\squared}$ for an excitation at $\sim$~\SI{450}{\nano\meter}~\cite{chenSizeWavelengthDependentTwoPhoton2017,puthenpurayilDeterminationAbsorptionCross2019}. We verify that the pulsed excitation experiments are performed in the linear regime where the number of excitons generated per pulse $N$ is kept below $\langle N\rangle\sim$~0.1. %
The emission intensity is stable over the course of hours (\myautoref[b]{fgr:exppow}) and no blinking is observed even at the smallest time bins (zoom-in view of \myautoref[b]{fgr:exppow}). %
Furthermore, only minimal spectral diffusion is observed at our resolution (\myautoref[d]{fgr:exppow}). %
A second-order correlation measurement performed on the bright exciton emission under pulsed excitation is shown in \myautoref[c]{fgr:exppow} and reveals highly antibunched emission. %

\section{Decay dynamics analysis}

\begin{figure*}
	\centering
	\sffamily
	\includegraphics{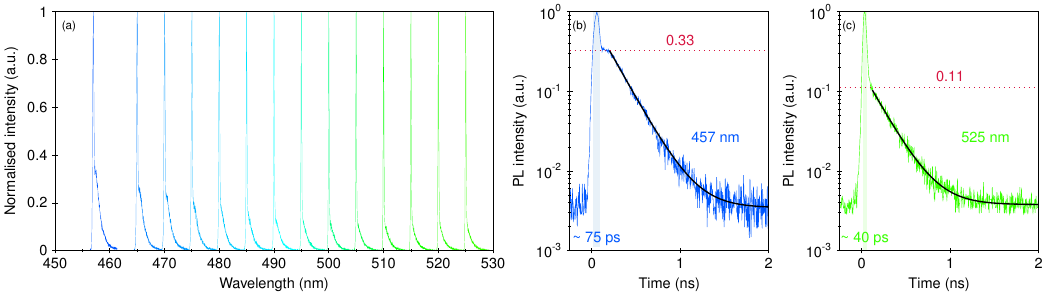}
	\caption{Instrument response function (IRF) of the time-correlated single-photon counting (TCSPC) setup. (a) IRF of the TCSPC setup as a function of wavelength. %
	(b,c) IRFs at (b) the excitation wavelength (\SI{457}{\nano\meter}) and (c) at the detection wavelength (\SI{525}{\nano\meter}) together with the key features: IRF peak width, tail/peak ratio and IRF tail decay.}
	\label{fgr:expIRF}
\end{figure*}
\subsection{Time-resolved data analysis}~\label{sec:IRF}

As we are concerned with measuring decay times in the order of \SI{100}{\pico\second} which is close to the nominal time-resolution of the single-photon detection setup, special attention needs to be given to the measurement of the instrument response function (IRF). 
Indeed, the finite width of the laser pulses as well as the physical and electronic nature of the single-photon detection process shape the temporal response of decay measurements. 
The IRF therefore exhibits (\myautoref{fgr:expIRF}) a characteristic shape with a sharp and narrow peak and a lower amplitude tail due to the diffusion of photocarriers generated in neutral regions near the depletion layer~\cite{kapustaAdvancedPhotonCounting2015}. %

To properly account for this response, we measure the IRF of our setup with a \SI{5}{\nano\meter} step from~\SIrange{460}{525}{\nano\meter} using the scattering of the pulsed laser excitation obtained by frequency-doubling a Ti:Sa laser (\SI{80}{\mega\hertz} with $\sim$~\SI{100}{\femto\second} pulses). The laser pulse is highly attenuated using the same filter used during our experiments and the count rate on the SPADs is kept similar to the single NC emission count rate (well below 1 photon per pulse). Interestingly, we find that the shape of the IRF is highly dependent on the detection wavelength even within this small window. %
\myautoref[a]{fgr:expIRF} shows the measured IRFs while in \myautoref[b,c]{fgr:expIRF}, we highlight the difference between the IRF measured at \SI{457}{\nano\meter} and \SI{525}{\nano\meter} together with the features of interest: tail/peak ratio, IRF width and characteristic decay time of the tail. The tail/peak ratio and IRF width show a similar decreasing trend with increasing wavelength from \num{0.33} to \num{0.11} and $\sim$~\SI{75}{\pico\second} to $\sim$~\SI{40}{\pico\second}, respectively. The characteristic decay time of the IRF tail is found to be nearly constant across the entire detection range at $\sim$~\SI{210}{\pico\second}. The characteristic IRF decay time is very similar to the decay times previously reported in the literature for single NC or ensembles at low temperature~\cite{fuNeutralChargedExciton2017, isarovRashbaEffectSingle2017, utzatCoherentSinglephotonEmission2019}, it is thus of utmost importance to properly take into account this response when fitting the time-resolved data.

\begin{figure*}
	\centering
	\includegraphics{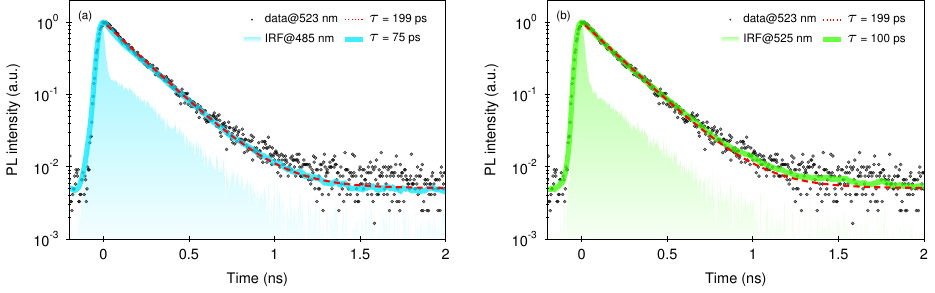}
	\caption{Influence of the IRF on the recovered decay parameters. Comparison of decay times obtained using different fit methods taking into account the IRF or not. (a,b) Full decay trace showing the raw data and fit without taking into account the IRF. %
	(a) Decay analysis with the IRF taken at \SI{485}{\nano\meter} together with the corresponding fit.
	(b) Decay analysis with the IRF taken at \SI{525}{\nano\meter} with the corresponding fit.
	}
	\label{fgr:compIRFs}
\end{figure*}

To analyse the decay curves, we perform a reconvolution fit using a maximum likelihood estimator to fit the decay curves with multi-exponential decays convoluted to the IRF as:
\begin{equation}\tag{S1}
I(t) = \sum_{i=1}^{N} \mathrm{IRF}(t) \otimes  A_i \exp\left( -\frac{t-t_0}{\tau_i}\right) + B
\end{equation}
where the IRF is measured within \SI{2.5}{\nano\meter} of the decay curve wavelength, the $A_i$ are the amplitudes of the exponential decays, $t_0$ is the zero-delay time, $\tau_i$ are the exponential decay times and $B$ is the correction for the decay background. %
For temperature-dependent decay measurements, decay curves are acquired as a function of temperature and each curve is fitted as a convolution of the IRF closest to the emission wavelength and a multi-exponential decay. This reconvolution technique taking into account the appropriate IRF thus yields reliable amplitudes for each component of the decay which cannot be obtained by fitting only the tail of decays (shown in \myautoref{fgr:compIRFs} in red). This allows us to monitor the temperature-dependence of the multi-exponential decay of single NCs via both the decay times and relative component intensity. 

To highlight the importance of the IRF, we show in \myautoref{fgr:compIRFs} a decay curve acquired at \SI{523.5}{\nano\meter} and several fit attempts. The lines in green, light blue and red show respectively the fit attempts using the IRF recorded at the closest wavelength of \SI{525}{\nano\meter}, the IRF recorded at \SI{485}{\nano\meter}, and, neglecting the IRF, only fitting the tail of the decay curve. Each of these fits converges with residuals that are evenly distributed around 0. As such, they could all seem to be reasonable but physically they are not. Each yields a different decay time shown in the figure legend. Interestingly, the IRF tail has a great impact on the decay time recovered from the fit procedure. When neglecting the IRF in the fit procedure, the decay time increases by a factor $\sim$2 and becomes close to the IRF tail decay time $\sim$~\SI{210}{\pico\second}. It is thus necessary, in order to accurately measure such short decay times, to take into account the IRF close to the emission wavelength particularly in this range as the detectors response can vary. %

We further verify the absence of additional unresolved decay times longer than the excitation laser repetition rate. To do so, we check the background level of photoluminescence decay curves against the background level obtained with no emitter, only due to the SPADs dark counts, during the same acquisition time. Both background levels being the same, we can exclude unresolved longer decay times.

\subsection{Model}~\label{sec:model}
\subsubsection{Rate equations}

\begin{figure}
	\includegraphics{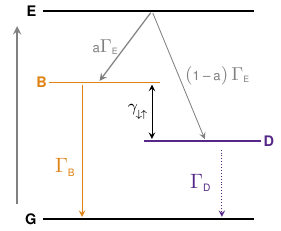}
	\caption{Schematic of the system comprised of a ground state (G), an excited state (E), a bright exciton state (B) and a dark exciton state (D) together with the rates involved in the model.}\label{fgr:themodel}
\end{figure}

The evolution of the system, shown in \myautoref{fgr:themodel}, is governed by rate equations that can be written in matrix form as
$$    \mathcal{\dot{P}} = \mathcal{A} \mathcal{P} \label{eq:mateq}$$ where $\mathcal{P} = \{P_i\}_{i\leq4}$ and $\mathcal{A}$ is the transition matrix given by

\begin{widetext}
\begin{equation*}
\begin{pmatrix}
\dot{p}_\text{G}\\
\dot{p}_\text{B} \\
\dot{p}_\text{D} \\
\dot{p}_\text{E}
\end{pmatrix}  = \begin{pmatrix}
0 							& 			\Gamma_\text{B} 									& 			\Gamma_\text{D} 								& 			0\\
0 				& -\left( \Gamma_\text{B} + \gamma_{\text{BD}}\right) 	& 			\gamma_{\text{DB}} 							& 		a \Gamma_\text{E}\\
0	& 		\gamma_{\text{BD}} 									& -\left( \Gamma_\text{D} + \gamma_{\text{DB}}\right) 	& \left(1-a\right) \Gamma_\text{E} \\
0 							& 				0 											& 					0 									& -\Gamma_{\text{E}}
\end{pmatrix}\begin{pmatrix}
p_\text{G}\\
p_\text{B} \\
p_\text{D}\\
p_\text{E}
\end{pmatrix}
\end{equation*}
\end{widetext}

At this point, the rate equations are general enough to allow for the scenarios that we will explore, i.e. one- and two-phonon transitions with two relative bright-dark orderings: bright over dark and bright under dark. This versatility is hidden in the transition rates $\gamma_{\text{BD}}$ and $\gamma_{\text{DB}}$.

The solution of the matrix equation is given by:
$$ \mathcal{P} =  \sum_{i=1}^N \alpha_i \vec{v}_i e^{- \left|\Gamma_i\right| t}$$ where $\alpha_i$ are constants derived from initial conditions, $\Gamma_i$ are the eigenvalues of $\mathcal{A}$ and $\vec{v}_i$ the corresponding eigenvectors. %
At $t=0$, we have $\mathcal{P}=(0,0,0,1)$, such that the $\alpha_i$ are given by the solution of ${\sum_i \alpha_i \vec{v}_i = \mathcal{P}(t=0)=(0,0,0,1)}$.

The total photoluminescence signal is given by
$$ S(t) = \textstyle\sum_i \eta_i \Gamma_i p_i = \sum_i A_i e^{-\left|\Gamma_i\right| t} $$
where $\eta_i$ is the quantum yield associated with state $i$ and the amplitudes read $ A_i = \textstyle \sum_j \eta_j \Gamma_j \alpha_i \vec{v}_i^{(j)} $
where $\vec{v}_i^{(j)}$ is the $j$-th component of eigenvector $\vec{v}_i$. The sum $i$ runs over $i=\text{B},\text{D},\text{E}$.

In our experiments, we only acquire the PL of the bright state given by 
$S_\text{B}(t) = \eta_\text{B} \Gamma_\text{B} p_\text{B} $ %
which in general is a tri-exponential decay of the form
$$A_{\text{short}} e^{-\Gamma_{\text{short}}t} + A_{\text{long}} e^{-\Gamma_{\text{long}}t} + A_{\text{rise}} e^{-\Gamma_{\text{rise}}t}$$
where $A_{\text{short}},A_{\text{long}}\geq 0$ and $A_{\text{rise}}\leq 0$.

In most cases, no rise time is observed (i.e. rise time constant is very short and/or the amplitude $A_{\text{rise}}$ is negligible) and the decays take the simpler form %
$$A_{\text{short}} e^{-\Gamma_{\text{short}}t} + A_{\text{long}} e^{-\Gamma_{\text{long}}t} $$

Using the reconvolution technique we can extract the values of the short and long decay times together with their relative amplitudes. We define the fractional intensity of the long component $I_{\text{rel},\text{long}}$ that will be used together with the decay times $\Gamma_\text{short}$ and $\Gamma_\text{long}$ for our analysis.

\begin{equation}
I_{\text{rel},\text{long}} = \left( \frac{A_{\text{long}}}{\Gamma_{\text{long}} }\right) / \left( \sum_i A_i/\Gamma_i \right)
\end{equation}

\begin{table*}[hbt]
	{\small
	\begin{tabular}{@{}lcccccc@{\hskip 15pt}} 
	\cmidrule[\heavyrulewidth]{2-3} \cmidrule[\heavyrulewidth](l){4-7}
	&%
	\multicolumn{2}{c}{\textbf{One-phonon process}}
	&%
	\multicolumn{4}{c}{\textbf{Two-phonon processes}}
	\\ 
	\cmidrule[\heavyrulewidth]{2-3} \cmidrule[\heavyrulewidth](l){4-7}
	 &%
	\BoDone  &%
	\BuDone  &%
	\BoDtwosum &%
	\BuDtwosum 
	&%
	\BoDtwodif  &%
	\BuDtwodif 
	\\ 
	 \addlinespace
	$\gamma_{\text{B}\rightarrow\text{D}}/\gamma_0$ &%
	$\left( N +1 \right)$ 	&%
	$N$  &%
	$\left(N_\text{1} +1 \right)\left(N_\text{2} +1 \right)$ &%
	$N_\text{1} N_\text{2}$ &%
	$N_\text{1}\left( N_\text{2} +1 \right)$ & %
	$N_\text{2}\left( N_\text{1} +1 \right)$   \\ \addlinespace
	$\gamma_{\text{D}\rightarrow\text{B}}/\gamma_0$ & %
	$N$ 		&%
	$\left( N +1 \right)$ & %
	$N_\text{1} N_\text{2}$ &%
	$\left(N_\text{1} +1 \right)\left(N_\text{2} +1 \right)$ &%
	$N_\text{2}\left( N_\text{1} +1 \right)$ & %
	$N_\text{1}\left( N_\text{2} +1 \right)$ \\ \addlinespace
	$\Gamma_{\text{short}} (T\rightarrow 0)$ &%
	$\Gamma_\text{B}$ + $\gamma_0$ &%
	$\Gamma_\text{B}$ &%
	$\Gamma_\text{B}$ + $\gamma_0$&%
	$\Gamma_\text{B}$ &%
	$\Gamma_\text{B}$ &%
	$\Gamma_\text{B} $ \\ \addlinespace
	$\Gamma_{\text{long}} (T\rightarrow 0)$\textcolor{blue}{$^*$} &%
	$\Gamma_\text{D}$ &%
	$\Gamma_\text{D}$ + $\gamma_0$ &%
	$\Gamma_\text{D}$&%
	$\Gamma_\text{D}$ + $\gamma_0$&%
	$\Gamma_{\text{D}}$ &%
	$\Gamma_\text{D}$ \\ \\ \addlinespace
	 \bottomrule\end{tabular} }
	 \caption{Four scenarios arising from two relative orderings and two phonon-driven mixing schemes. \textcolor{blue}{$^*$}Since the relative intensity of this component is less than a few \% at the lowest temperatures this limit is not used as a constraint.}
	 \label{tbl:scenarios}
	\end{table*}

\subsubsection{Simulations}~\label{sec:simparams}

Transitions between exciton sublevels are modelled as phonon-assisted mechanisms. Here, we consider both first-order and second-order phonon-driven transitions with the two possible level orderings: bright state above or below the dark state. %
The corresponding bright-dark transfer rates are given in the main text and summarised in~\autoref{tbl:scenarios}. %

As equations are quite intricated and we have quite a lot of parameters, we would like to rely on more than just non-linear optimisation as it is often difficult to assess the relative quality of fits solely based on residuals or goodness of fit measurements with this high number of parameters. %
In~\autoref{tbl:scenarios}, the low- and high-temperature limits of the short and long decay components are given and used to constrain the model parameters $\Gamma_\text{B}$ and $\Gamma_\text{D}$. %

Let us emphasize that $a$ is set to $0.75$ to reflect the exciton fine structure of our material consisting of a bright triplet and a dark singlet. In fact, this parameter also controls the maximum decay-component relative intensity attainable under strong magnetic field, i.e. $I_{\text{rel,long}} \left(B \gg \Delta_0/\mu_B \Delta_\text{BD} g \right) \rightarrow 1 - a$. %
The set value of $a=0.75$ shows very good agreement with measurements on ensembles of \ch{CsPbBr3} NCs at magnetic fields up to \SI{30}{\tesla} where $I_\text{rel,long}(\SI{30}{\tesla})\sim 0.25$~\cite{cannesonNegativelyChargedDark2017}. %

\subsubsection{Conversion efficiency}~\label{sec:conversionefficiency}

\begin{figure}[htbp]
    \centering
    \includegraphics{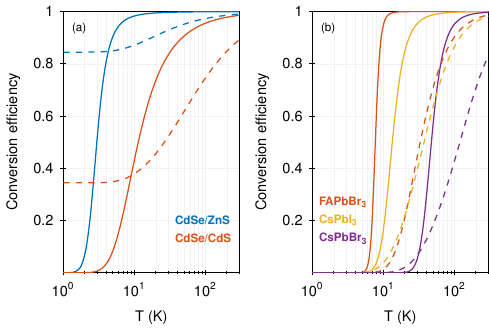}
    \caption{Conversion efficiencies between bright and dark excitons in several materials. Dark-to-bright (solid) and bright-to-dark (dashed) exciton conversion efficiency (a) for CdSe-based NCs with one-phonon bright-dark population transfer and (b) for LHP NCs with two-phonon bright-dark population transfer. Transition rates used in this calculation for CsPbBr\textsubscript{3} are $\gamma_0=\SI{10}{\nano\second^{-1}}$, $\Gamma_\text{B}=\SI{15}{\nano\second^{-1}}$, $\Gamma_\text{D}=\SI{0.12}{\nano\second^{-1}}$ with phonon energies $E_{\text{ph},1}=\SI{6}{\milli\electronvolt}$, $E_{\text{ph},2}=\SI{19}{\milli\electronvolt}$ and $a=0.75$ while rates for other materials are extracted from refs.~\cite{biadalaDirectObservationTwo2009,werschlerCouplingExcitonsDiscrete2016,tamaratGroundExcitonState2019,tamaratDarkExcitonGround2020}.}
    \label{fig:conveff}
\end{figure}

In the case of a dark ground state, we can define the conversion efficiency from bright to dark, resp. dark to bright, exciton as $\gamma_\downarrow/(\Gamma_\text{B} + \gamma_\downarrow)$, resp. $\gamma_\uparrow/(\Gamma_\text{D} + \gamma_\uparrow)$. In \myautoref{fig:conveff} we show the conversion efficiencies from bright to dark and dark to bright exciton for both CdSe-based NCs and LHP NCs. 
For CdSe-ZnS NCs, the conversion efficiency from bright to dark exciton is larger than 80\% at the lowest temperatures and increases with temperature. Similarly, for CdSe-CdS NCs, more than 35\% of the bright exciton population is converted to the dark state at the lowest temperatures and 50\% at $\sim$~\SI{26}{\kelvin}. %
Conversely, the conversion efficiency from dark to bright exciton is zero at the lowest temperatures and reaches 50\% at \SI{3}{\kelvin} and \SI{11}{\kelvin} respectively. %
This contrasts with LHPs, where the conversion efficiency from bright to dark exciton as well as dark to bright exciton are both zero at the lowest temperatures. The dark to bright conversion efficiency reaches 50\% at 8, 13, 16 and \SI{54}{\kelvin}, respectively for FAPbBr\textsubscript{3}, CsPbI\textsubscript{3}, FAPbI\textsubscript{3} and CsPbBr\textsubscript{3} and the bright to dark conversion efficiency reaches 50\% at 33, 37, 73 and \SI{162}{\kelvin}. %
This highlights the weak efficiency of the bright/dark population conversion in LHP NCs which stems from the second-order nature of the bright/dark transitions in LHPs and more precisely from the vanishing upward and downward transition rates at the lowest temperatures coupled with favourable transition rates. %

\subsubsection{Integrated intensities}~\label{sec:intint}

Within the system comprised of a dark and a bright state, we can determine the total emitted intensity $I_i$ of a state $i$ by integrating its population as $I_i = \eta_i \int_0^\infty \Gamma_i p_i dt$. %
The stationary solution can instead be obtained by directly integrating the system of equations, from which we obtain the following expressions for the integrated intensities of the bright and dark states for a dark ground state: %
\begin{align*}
I_\text{B} &= \eta_\text{B} \Gamma_\text{B} \frac{p_\text{B}(0) \left(\Gamma_\text{D} + \gamma_\uparrow\right) + p_\text{D}(0) \gamma_\uparrow}{\left(\Gamma_\text{B} +\gamma_\downarrow\right)\left(\Gamma_\text{D} + \gamma_\uparrow\right) - \gamma_\uparrow\gamma_\downarrow}\\
I_\text{D} &= \eta_\text{D} \Gamma_\text{D} \frac{p_\text{D}(0) \left(\Gamma_\text{B} + \gamma_\downarrow\right) + p_\text{B}(0) \gamma_\downarrow}{\left(\Gamma_\text{B} +\gamma_\downarrow\right)\left(\Gamma_\text{D} + \gamma_\uparrow\right) - \gamma_\uparrow\gamma_\downarrow}
\end{align*}
where $p_B(0)=a$ and $p_D(0)=1-a$.
The ratio of intensity $I_\text{B}/I_\text{D}$ therefore reads:
\begin{equation}
\frac{I_\text{B}}{I_\text{D}} = \frac{\eta_\text{B} \Gamma_\text{B}}{\eta_\text{D} \Gamma_\text{D}} \frac{p_\text{B}(0) \left(\Gamma_\text{D} + \gamma_\uparrow\right) + p_\text{D}(0) \gamma_\uparrow}{p_\text{D}(0) \left(\Gamma_\text{B} + \gamma_\downarrow\right) + p_\text{B}(0) \gamma_\downarrow}
\end{equation}

For one-phonon transitions and for the two-phonon sum process, the ratio of intensities as $T\rightarrow 0$ simplifies to
\begin{equation}
	\frac{I_\text{B}}{I_\text{D}}\Bigg|_\text{1ph} = \frac{I_\text{B}}{I_\text{D}}\Bigg|_\text{2ph,sum} = \frac{\eta_\text{B}}{\eta_\text{D}} \frac{a}{(1-a) + \gamma_0/\Gamma_\text{B}}
\end{equation}
The low temperature intensity ratio $I_\text{B}/I_\text{D}$ therefore depends not only on $a$ but also on $\gamma_0/\Gamma_\text{B}$.

In the case of the two-phonon difference process, the ratio of intensities as $T\rightarrow 0$ simplifies to
\begin{equation}
	\frac{I_\text{B}}{I_\text{D}}\Bigg|_\text{2ph,Raman} = \frac{\eta_\text{B}}{\eta_\text{D}} \frac{p_\text{B}(0)}{p_\text{D}(0)} = \frac{\eta_\text{B}}{\eta_\text{D}} \frac{a}{1-a}
 \label{eq:IBID}
\end{equation}
where $a=0.75$, such that regardless of the actual rates involved in the system, the ratio of intensity between the bright and dark states is given by $3\eta_\text{B}/\eta_\text{D}$. %
This is in stark contrast with the case of one-phonon transitions and the two-phonon sum process, where the ratio of intensities depends on the actual rates involved in the system via $\gamma_0/\Gamma_\text{B}$. %

In the case of a bright ground state, the relative intensity as $T\rightarrow 0$ reads:
\begin{equation}
	\frac{I_\text{B}}{I_\text{D}}\Bigg|_\text{1ph} = \frac{I_\text{B}}{I_\text{D}}\Bigg|_\text{2ph,sum} = \frac{\eta_\text{B}}{\eta_\text{D}} \frac{a + \gamma_0/\Gamma_\text{D}}{(1-a) }
\end{equation}
The intensity ratio depends not only on $a$ but also on the ratio $\gamma_0/\Gamma_\text{D}$.

For the second-order difference process, we find that the low-temperature relative intensities do not depend on the actual ordering of the exciton sublevels:
\begin{equation}
	\frac{I_\text{B}}{I_\text{D}}\Bigg|_\text{2ph,Raman} = \frac{\eta_\text{B}}{\eta_\text{D}} \frac{p_\text{B}(0)}{p_\text{D}(0)} = \frac{\eta_\text{B}}{\eta_\text{D}} \frac{a}{1-a}
\end{equation}

\subsection{Simulation procedure}~\label{sec:procedure}

Here we detail the procedure used to obtain the characteristic temperature $T_c$ in the phase sub-space where all experimental requirements are met. %

\begin{figure}[ht]
	\includegraphics{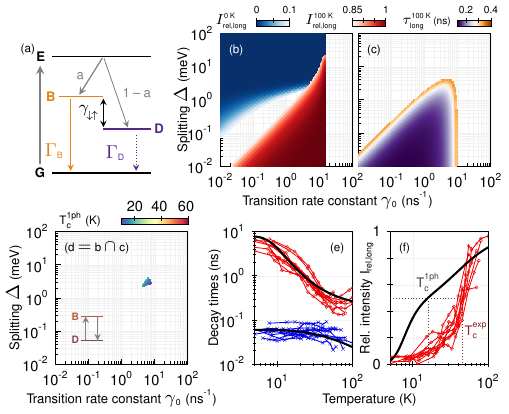}
	\caption{Phase space mapping of first-order phonon transitions. (a) Schematic of the system comprised of a ground state (G), an excited state (E), a bright exciton state (B) and a dark exciton state (D) and the coupling rates of the model. (b,c) Particular features of interest obtained from simulations of the decays in the $(\gamma_0,\, \Delta_\text{BD})$ space. (b) Intensity of the long-decay component at 0 and \SI{100}{\kelvin}. (c) Long decay time at \SI{100}{\kelvin}. (d) Characteristic temperature $T_c^{\text{1ph}}$ in the phase subspace defined by the intersection of the (b) and (c) subspaces. (e, f) Experimental decay dynamics with (e) the long and short decay times and (f) the long decay component fractional intensity together with a fit using parameters in (d).}
	\label{fig:BoD1phmapsSI}
\end{figure}

We calculate the long-decay-time intensity at in the low- and high-temperature limits, $I_\text{rel,long}^\text{0 K}$ and $I_\text{rel,long}^\text{100 K}$, as well as the long decay time in the high-temperature limit $\tau_{\text{long}}^{100 K}$. For the sake of clarity, we restrict the phase sub-space displayed to the one matching our experimental results, i.e. we only show the region of phase space where the long-decay-time relative intensity is lower than 0.1 at \SI{0}{\kelvin} and higher than 0.85 at \SI{100}{\kelvin} (\myautoref[b]{fig:BoD1phmapsSI}), and where the long decay time is comprised between \SI{150}{\pico\second} and \SI{350}{\pico\second} at \SI{100}{\kelvin} (\myautoref[c]{fig:BoD1phmapsSI}).
	
The phase subspace in \myautoref[d]{fig:BoD1phmapsSI} (same as \myautoref[b]{Mfig:BoDNphmaps}) is then defined as the intersection of the three subspaces related to the long decay fractional intensity at low and high temperature (\myautoref[b]{fig:BoD1phmapsSI}) and to the long decay time at high temperature (\myautoref[c]{fig:BoD1phmapsSI}). %
Applying the same procedure for the opposite fine structure ordering, with a bright exciton below the dark exciton, yields an empty intersection.

\begin{figure}[h]
    \centering
	\includegraphics{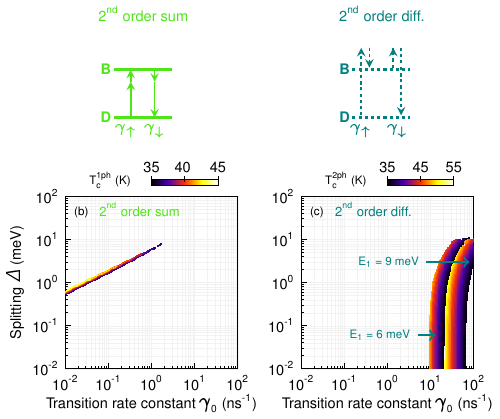}
    \caption{Additional phase space mapping of second-order processes. (a) Sum-process with $\hbar\omega_2 = 9 \hbar\omega_1$ such that $\hbar\omega_1 + \hbar\omega_2 = \Delta_\text{BD}$. (b) Difference-process with varying the lower phonon energy $E_1 = \SI{6}{\milli\electronvolt}$ and $E_1=\SI{9}{\milli\electronvolt}$.}
    \label{fig:CoT_extra}
\end{figure}

For second-order transitions, the same procedure is applied to obtain \myautoref[c]{Mfig:BoDNphmaps}. This procedure was also applied for the opposite fine structure ordering, with the bright exciton state below the dark exciton state, for which we find that only the two-phonon difference model yields a non-empty intersection.

As discussed in the main text, for both second-order processes the energy of one of the phonon modes can further be varied while ensuring $\hbar\omega_1 + \hbar\omega_2 = \Delta_\text{BD}$ and $\hbar\omega_2-\hbar\omega_1=\Delta_\text{BD}$. \myautoref{fig:CoT_extra} shows that the obtained phase spaces are similar to the ones obtained in \myautoref[b,c]{Mfig:BoDNphmaps}. For the second-order difference process, the obtained phase space is slightly shifted towards larger $\gamma_0$ as the lowest phonon energy is increased from 3 to 6 and \SI{9}{\milli\electronvolt}, matching the optical phonon modes observed as replica in the spectrum.

\section{Additional data}

\subsection{Bandgap thermal dependence}~\label{sec:allenergies}

\begin{figure}[htbp]
    \centering
	\includegraphics{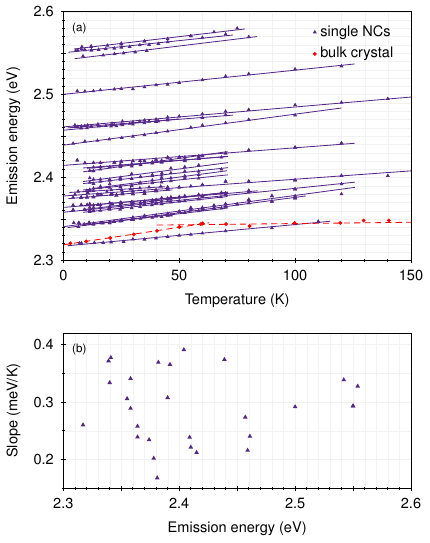}
    \caption{Temperature-dependence of the emission energy. (a) Bright triplet emission energy as a function of temperature. In red, we reproduced the dependence recorded on bulk single crystals in ref.~\cite{guoDynamicEmissionStokes2019}. (b) Emission energy dependence of the slope of $E(T)$ deduced from (a).}
    \label{fig:EfT}
\end{figure}

In the investigated temperature range, the emission energy of single \ch{CsPbBr3} NCs increases linearly with temperature (\myautoref[a]{fig:EfT}) with an average slope of $\sim$~\SI{0.3}{\milli\electronvolt\per\kelvin} (\myautoref[b]{fig:EfT}), slightly smaller than the low temperature slope measured on single bulk crystals~\cite{guoDynamicEmissionStokes2019}. %

\subsection{Decay time statistics}~\label{sec:decaystats}

\begin{figure}[h]
\centering
\includegraphics{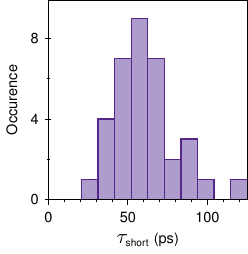}
\caption{Short decay time statistics at \SI{4.5}{\kelvin}.}
\label{fig:decaystats}
\end{figure}

As discussed in the main text, the PL decay of single \ch{CsPbBr3} NCs is bi-exponential at all temperatures with a predominant sub-\SI{100}{\pico\second} decay channel at the lowest temperatures. In \myautoref{fig:decaystats}, we show the distribution of this short decay time at \SI{4.5}{\kelvin} measured across $\sim$~35 NCs.

\subsection{Temperature-dependent photoluminescence}

\begin{figure}[h]
    \centering
	\includegraphics{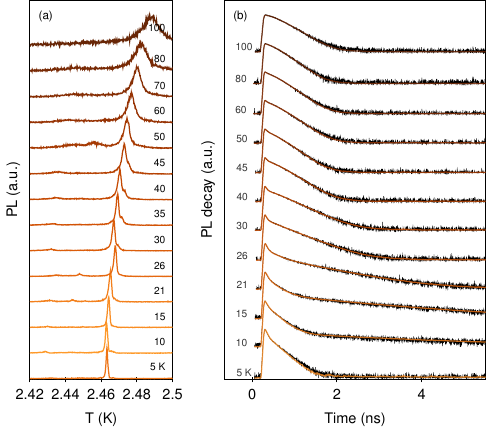}
    \caption{Temperature-dependent photoluminescence. (a) Steady-state PL spectra as a function of temperature. (b) Time-resolved PL as a function of temperature with bi-exponential fits.}
    \label{fig:dataset}
\end{figure}

\myautoref{fig:dataset} shows an example of the full PL temperature-dependence recorded with the steady-state PL in \myautoref[a]{fig:dataset} and the fitted time-resolved PL in \myautoref[b]{fig:dataset}.

\subsection{Dark exciton state}~\label{sec:darkexp}

We performed magneto-optical spectroscopy on single \ch{CsPbBr3} NCs that reveals the emergence of a dark singlet state below the bright triplet. %
Based on prior studies~\cite{fuNeutralChargedExciton2017,choExcitonPhononTrion2022,amaraSpectralFingerprintQuantum2023,tamaratUniversalScalingLaws2023} and comparing finite field measurements to zero-field measurements, we can safely identify the trion and optical phonon replica.
For \myautoref[a]{fig:PLfBB}, at zero field we can identify the three lowest energy optical phonon replica. Based on the bright triplet emission energy, the trion emission is expected around $\sim~\SI{30}{\milli\electronvolt}$ and is not observed. At finite field, a weak additional peak emerges below the bright triplet at $\sim$~\SI{14}{\milli\electronvolt}, close to an optical phonon replica, and is attributed to the brightened dark singlet. %
For \myautoref[b]{fig:PLfBB}, we observe the weak emergence of two peaks below the bright triplet in addition to previously identified optical phonon replica. Based on the emission energy of the bright triplet, the trion emission is around $\sim~\SI{23}{\milli\electronvolt}$ and we find a peak at $\sim~\SI{21.5}{\milli\electronvolt}$. %
It is therefore attributed to the trion whose emission emerged after prolonged exposure to the excitation laser. %
The second peak appears at finite magnetic field at $\sim~\SI{12}{\milli\electronvolt}$ with a lower energy than the previously identified low-energy optical phonon replica and is attributed to the dark singlet. %

\begin{figure}[h]
    \centering
	\includegraphics{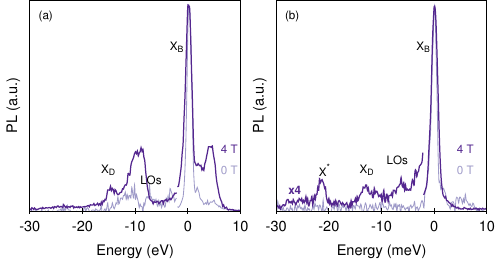}
    \caption{Magneto-optical spectroscopy of individual \ch{CsPbBr3} NCs at \SI{4.5}{\kelvin}. (a) NC emitting at \SI{2.48}{\electronvolt} and (b) NC emitting at \SI{2.42}{\electronvolt}.}
    \label{fig:PLfBB}
\end{figure}

In both cases, an additional peak appears at finite magnetic field at a lower energy than the previously identified low-energy optical phonon replica, with an energy splitting well above the bulk bright-dark splitting of $\sim~\SI{3.5}{\milli\electronvolt}$~\cite{tamaratUniversalScalingLaws2023}. %
Using the recently determined scaling law, we can compare our results to those obtained on weakly confined \ch{CsPbI3} NCs~\cite{tamaratUniversalScalingLaws2023}. Using ${E_g=\SI{2.342}{\electronvolt}}$ and ${R_y=\SI{33}{\milli\electronvolt}}$~\cite{yangImpactHalideCage2017}, we find normalised energies ${(E_X-E_g)/R_y = 2.36\ \text{and}\ 4.2}$ with corresponding bright-dark splittings ${\Delta_\text{BD}/\left(R_y^2/E_g\right) = 27\ \text{and}\ 30}$ which fall within the range of values obtained for \ch{CsPbI3} NCs in a similar confinement regime and therefore provides further support for these identifications. %

\section*{References}

\bibliographystyle{achemso}
\bibliography{bdmanu}